\documentclass[prb,twocolumn,floatfix,superscriptaddress,amsmath,aps,longbibliography]{revtex4-2}
\pdfoutput=1
\usepackage[T1]{fontenc}\usepackage[latin1]{inputenc}
\usepackage{dcolumn,graphicx,color,booktabs,microtype,afterpage} %\graphicspath{{fig/}}
\usepackage[charter,greekuppercase=italicized]{mathdesign}\usepackage{sidecap}
\usepackage{amsmath}
\usepackage[colorlinks,plainpages=false,linkcolor=blue,urlcolor=blue,citecolor=blue,pdfpagemode=UseNone,pdfstartview=FitBH]{hyperref}
\graphicspath{{Figures/}}
\newcommand{\cusnoh}{CuSn(OH)$_6$}
\newcommand{\cusnod}{CuSn(OD)$_6$}

\begin{document}
	%
	%\preprint{\textit{Preprint: \today, \now. For internal use only, do not distribute.}}%\linenumbers
	%
	
\title{Spin Liquid Mimicry in the Hydroxide Double Perovskite \cusnod\ Induced by Correlated Proton Disorder}
	
\author{Anton\,A.\,Kulbakov}
\affiliation{Institut f{\"u}r Festk{\"o}rper- und Materialphysik, Technische Universit{\"a}t Dresden, 01062 Dresden, Germany}
	
\author{Ellen\,H\"au\ss ler}
\affiliation{Fakult{\"a}t f{\"u}r Chemie und Lebensmittelchemie, Technische Universit{\"a}t Dresden, 01062 Dresden, Germany}
	
\author{Kaushick~K.~Parui}
\author{Nikolai~S.~Pavlovskii}
\author{Aswathi~Mannathanath~Chakkingal}
\author{Sergey\,A.\,Granovsky}
%\email{sergey.granovsky@tu-dresden.de}
\affiliation{Institut f{\"u}r Festk{\"o}rper- und Materialphysik, Technische Universit{\"a}t Dresden, 01062 Dresden, Germany}
	
\author{Sebastian\,Ga\ss}
%\email{S.Gass@ifw-dresden.de}
\affiliation{Institute for Solid State Research, Leibniz IFW Dresden, 01069 Dresden, Germany}

\author{Laura\,Teresa {Corredor Boh{\'o}rquez}}
%\email{l.t.corredor.bohorquez@ifw-dresden.de}
\altaffiliation[Current affiliation: ]{Fakult{\"a}t f{\"u}r Physik, Technische Universit{\"a}t Dortmund, 44227 Dortmund, Germany}
\affiliation{Institute for Solid State Research, Leibniz IFW Dresden, 01069 Dresden, Germany}
 
\author{Anja\,U.\,B.\,Wolter}
%\email{a.wolter@ifw-dresden.de}
\affiliation{Institute for Solid State Research, Leibniz IFW Dresden, 01069 Dresden, Germany}
\affiliation{W\"urzburg-Dresden Cluster of Excellence on Complexity and Topology in Quantum Matter--ct.qmat, TU Dresden, 01062 Dresden, Germany}

\author{Sergei A.\,Zvyagin}
%\email{s.zvyagin@hzdr.de}
\affiliation{Dresden High Magnetic Field Laboratory (HLD-EMFL), Helmholtz-Zentrum Dresden-Rossendorf (HZDR), 01328 Dresden, Germany}

\author{Yurii\,V.\,Skourski}
\affiliation{Dresden High Magnetic Field Laboratory (HLD-EMFL), Helmholtz-Zentrum Dresden-Rossendorf (HZDR), 01328 Dresden, Germany}

\author{Vladimir\,Yu.\,Pomjakushin }
\affiliation{Laboratory\,for\,Neutron\,Scattering\,and\,Imaging\,(LNS),\,Center\,for\,Neutron\,and\,Muon\,Sciences\,(CNM),\,PSI,\,CH-5232\,Villigen,\,Switzerland}	

\author{In\'es Puente-Orench}
\affiliation{Instituto de Nanociencia y Materiales de Arag{\'o}n (INMA), CSIC-Universidad de Zaragoza, Zaragoza 50009, Spain}
\affiliation{Institut Laue-Langevin, 71 Avenue des Martyrs, CS 20156, CEDEX 9, 38042 Grenoble, France}
	
\author{Darren C.\,Peets}
\affiliation{Institut f{\"u}r Festk{\"o}rper- und Materialphysik, Technische Universit{\"a}t Dresden, 01062 Dresden, Germany}
	
\author{Thomas Doert}\email[Corresponding author: ]{thomas.doert@tu-dresden.de}
\affiliation{Fakult{\"a}t f{\"u}r Chemie und Lebensmittelchemie, Technische Universit{\"a}t Dresden, 01062 Dresden, Germany}
	
\author{Dmytro\,S.\,Inosov }\email[Corresponding author: ]{dmytro.inosov@tu-dresden.de}
\affiliation{Institut f{\"u}r Festk{\"o}rper- und Materialphysik, Technische Universit{\"a}t Dresden, 01062 Dresden, Germany}
\affiliation{W\"urzburg-Dresden Cluster of Excellence on Complexity and Topology in Quantum Matter--ct.qmat, TU Dresden, 01062 Dresden, Germany}
	
\begin{abstract}

The face-centered-cubic lattice is composed of edge-sharing tetrahedra, making it a leading candidate host for strongly frustrated magnetism, but relatively few face-centered frustrated materials have been investigated.  In the hydroxide double perovskite \cusnoh, magnetic frustration of the Cu$^{2+}$ quantum spins is partially relieved by strong Jahn-Teller distortions. Nevertheless, the system shows no signs of long-range magnetic order down to 45\,mK and instead exhibits broad thermodynamic anomalies in specific heat and magnetization, indicating short-range dynamical spin correlations\,---\,a behavior typical of quantum spin liquids. We propose that such an unusual robustness of the spin-liquid-like state is a combined effect of quantum fluctuations of the quantum spins $S=\frac{1}{2}$, residual frustration on the highly distorted face-centered Cu$^{2+}$ sublattice, and correlated proton disorder. Similar to the disorder-induced spin-liquid mimicry in YbMgGaO$_4$ and herbertsmithite, proton disorder destabilizes the long-range magnetic order by introducing randomness into the magnetic exchange interaction network. However, unlike the quenched substitutional disorder on the magnetic sublattice, which is difficult to control, proton disorder can in principle be tuned through pressure-driven proton ordering transitions. This opens up the prospect of tuning the degree of disorder in a magnetic system to better understand its influence on the magnetic ground state.

\end{abstract}
	
\maketitle

\section{Introduction}
A quantum spin liquid (QSL), proposed by Anderson in 1973\,\cite{Anderson_1973}, is an exotic state of magnetic matter that exhibits liquid-like behavior with strong quantum fluctuations and avoids long-range magnetic order even at absolute zero temperature\,\cite{Savary_2016,Balents_2010,Ramirez_2025}. QSLs are both fundamentally intriguing and a promising platform for quantum computing, and extensive theoretical and experimental efforts have been made to classify different types of QSLs and to find their material realizations\,\cite{Mei_2017,Lu_2024,Miksch_2021,Feng_2017,Yamashita_2011,Liu_2022,Fujihala_2020,Barkeshli_2013,Liu_2022a}. Avoiding long-range order despite strong interactions among the magnetic moments is generally accomplished through frustration, whereby exchange interactions compete with each other, preventing the system from readily selecting a unique ground state.  %Magnetic frustration is likely to be a key ingredient in obtaining a QSL ground state. 
Most spin liquids are highly fragile, as their stability is confined to a tiny region of parameter space. In the best-known candidate spin-liquid systems, minute lattice distortions or other deviations from the idealized spin model typically stabilize spin order at sufficiently low temperatures\,\cite{Kulbakov_2021, Avdoshenko_2022,Xie_2023}. 

The identification of spin-liquid states is challenging due to the lack of a discernible order parameter. An absence of magnetic order can be attributed to various factors, including quenched substitutional disorder or site intermixing. These phenomena show comparable signatures to spin liquids in thermodynamic measurements. This particular form of behavior, in which a QSL-like state is induced by quenched atomic disorder, has been termed ``spin-liquid mimicry'' to distinguish it from the true spin-liquid ground states, which persist even within an idealized, perfectly spatially periodic system~\cite{Zhu_2017}. Notable examples of such behavior are found in the mineral herbertsmithite ZnCu$_3$(OH)$_6$Cl$_2$~\cite{Freedman_2010,Zorko_2017,Smaha_2020,Huang_2021,Imai_2011,Mendels_2007,Helton_2007,Misguich_2007, Chitra_2008,Fu_2015,Han_2016,Khuntia_2020, Vries_2008,Norman_2016} and the triangular-lattice compound YbMgGaO$_4$~\cite{Li_2016,Li_2019,Zhu_2017}.

%Herbertsmithite, ZnCu$_3$(OH)$_6$Cl$_2$, is a natural mineral in which $S = 1/2$ Cu$^{2+}$ ions form a nearly ideal kagome lattice. The material, however, exhibits significant substitutional disorder: Zn$^{2+}$ occupies some Cu$^{2+}$ sites within the kagome layers, while Cu$^{2+}$ is found on some Zn$^{2+}$ interplane sites~\cite{Freedman_2010,Zorko_2017,Smaha_2020,Huang_2021,Imai_2011}. Such disorder can lead to spin-liquid mimicry. No long-range magnetic order is observed down to 50~mK despite strong antiferromagnetic interactions~\cite{Mendels_2007,Helton_2007,Misguich_2007, Chitra_2008}; however, the nature of the spin-liquid ground state remains under debate. Neutron and NMR data provide conflicting evidence. Some studies report a gapped~\cite{Imai_2011,Fu_2015,Han_2016} ground state while others suggest a gapless~\cite{Helton_2007,Khuntia_2020, Vries_2008} one. 

%YbMgGaO$_4$ is a synthetic compound where Yb$^{3+}$ with effective spin $1/2$ forms a two-dimensional triangular lattice~\cite{Li_2016}. It shows no magnetic order down to 50~mK. $\mu$SR measurements have revealed no spin freezing, indicating persistent spin dynamics compatible with a U(1)-type QSL state. Inelastic neutron scattering further revealed a gapless continuum of magnetic excitations at 60~mK~\cite{Li_2019}. However, random site mixing between Mg$^{2+}$ and Ga$^{3+}$ leads to a locally disordered crystalline environment around Yb$^{3+}$ ions, resulting in spatially varying exchange interactions~\cite{Zhu_2017}. This structural randomness mimics spin-liquid behavior in YbMgGaO$_4$.

\begin{figure}[b]
		\includegraphics[width=1.0\linewidth]{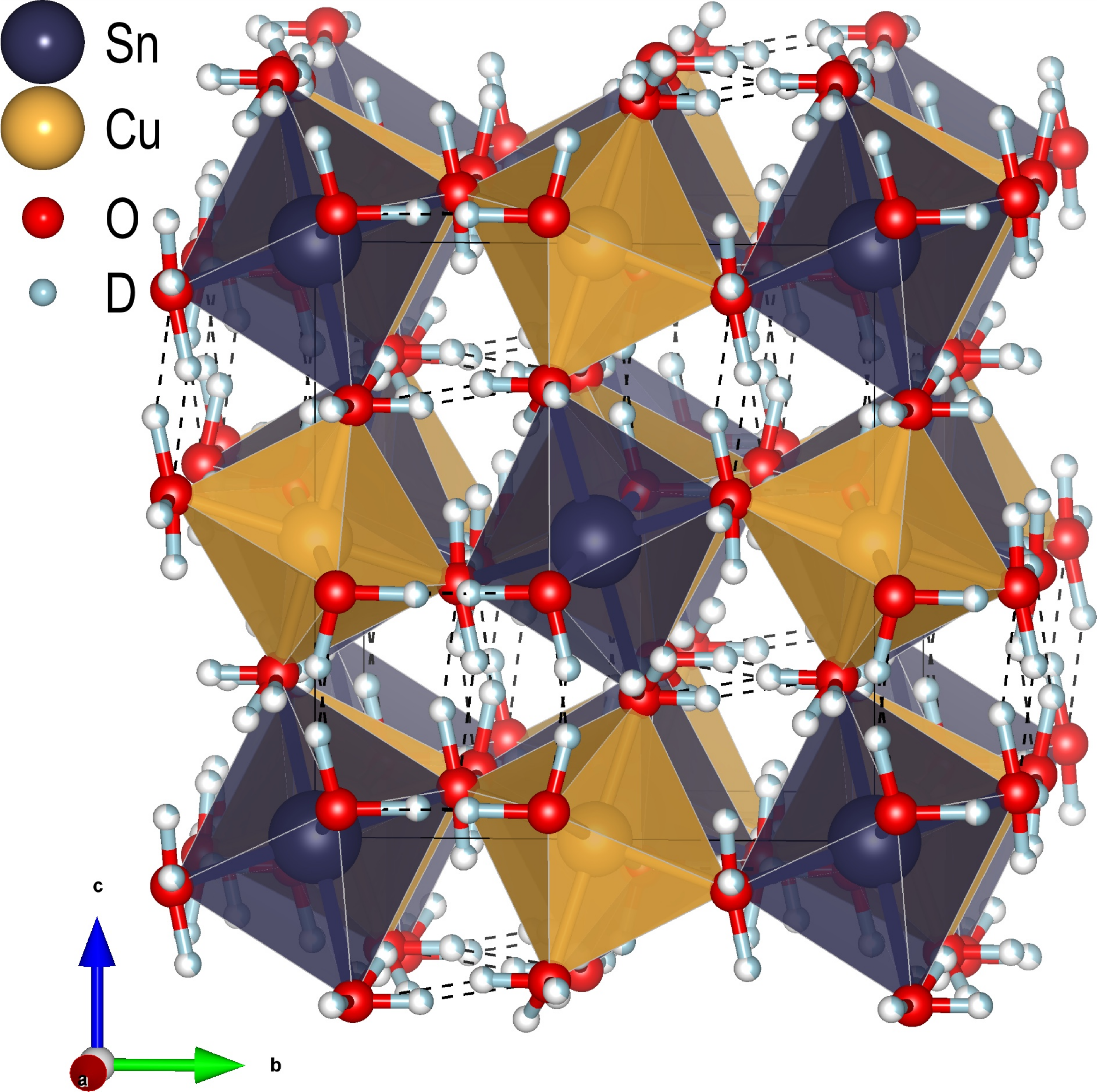}
		\caption{Crystal structure of \cusnod\ at 6\,K with partially occupied deuterium positions. The visualization was done in {\sc Vesta}\,\cite{VESTA}.}
		\label{fig:struct}
	\end{figure}

Hydroxide perovskites with correlated proton disorder~\cite{Welch_2025,LafuenteYang15,Kampf2024,Basciano98} appear promising in this respect, because their hydrogen sites obey analogies of the ice rules known from water ice and spin ices, and it is expected that this correlated proton disorder can be tuned by hydrostatic pressure~\cite{Kulbakov_2025} via a sequence of proton ordering transitions, similar to those known in conventional water ice~\cite{BernalFowler33,Pauling35,Malenkov09}. In this paper we focus on \cusnoh, a hydroxide double perovskite crystallizing in the orthorhombic $Pnnn$ space group~\cite{Kulbakov_2025}. The crystal structure of synthetic deuterated mushistonite, which we refined previously~\cite{Kulbakov_2025} from a combination of neutron and x-ray diffraction, is shown in Fig.\,\ref{fig:struct}. It consists of alternating [Cu$^{2+}$(OD)$_6$] and [Sn$^{4+}$(OD)$_6$] octahedra, with significant Jahn-Teller distortion of the Cu environment. 
  
In this work, we investigate the low-temperature magnetic properties of mushistonite as the first example of a material exhibiting spin-liquid mimicry induced by proton disorder. We investigate its magnetic behavior through heat capacity, magnetic susceptibility, and powder neutron diffraction measurements, and find that it is likely a uniquely suitable material for continuously tuning spin-liquid mimicry.

	\section{Experimental details}
\subsection{Sample preparation} \label{SubSec:synthesis}
	
    Phase-pure powder samples of \cusnoh\ were synthesized by co-precipitation from aqueous solutions of CuCl$_2\cdot2$H$_2$O and Na$_2$Sn(OH)$_6$ at room temperature. Equimolar amounts of the starting materials were ground together before dissolving them in water with concentration 0.16\,mol/L. After six days of vigorous stirring the precipitate was filtered, washed with ethanol, and dried in a vacuum drying chamber. The deuterated version, \cusnod, was synthesized by the same procedure from a solution in D$_{2}$O. The resulting pale blue powder contains intergrown cubic crystallites less than 1\,$\mu$m in size.
	
	\subsection{Specific heat and magnetization}
	\label{SubSec:TD}

	\begin{figure*}
		\includegraphics[width=\linewidth]{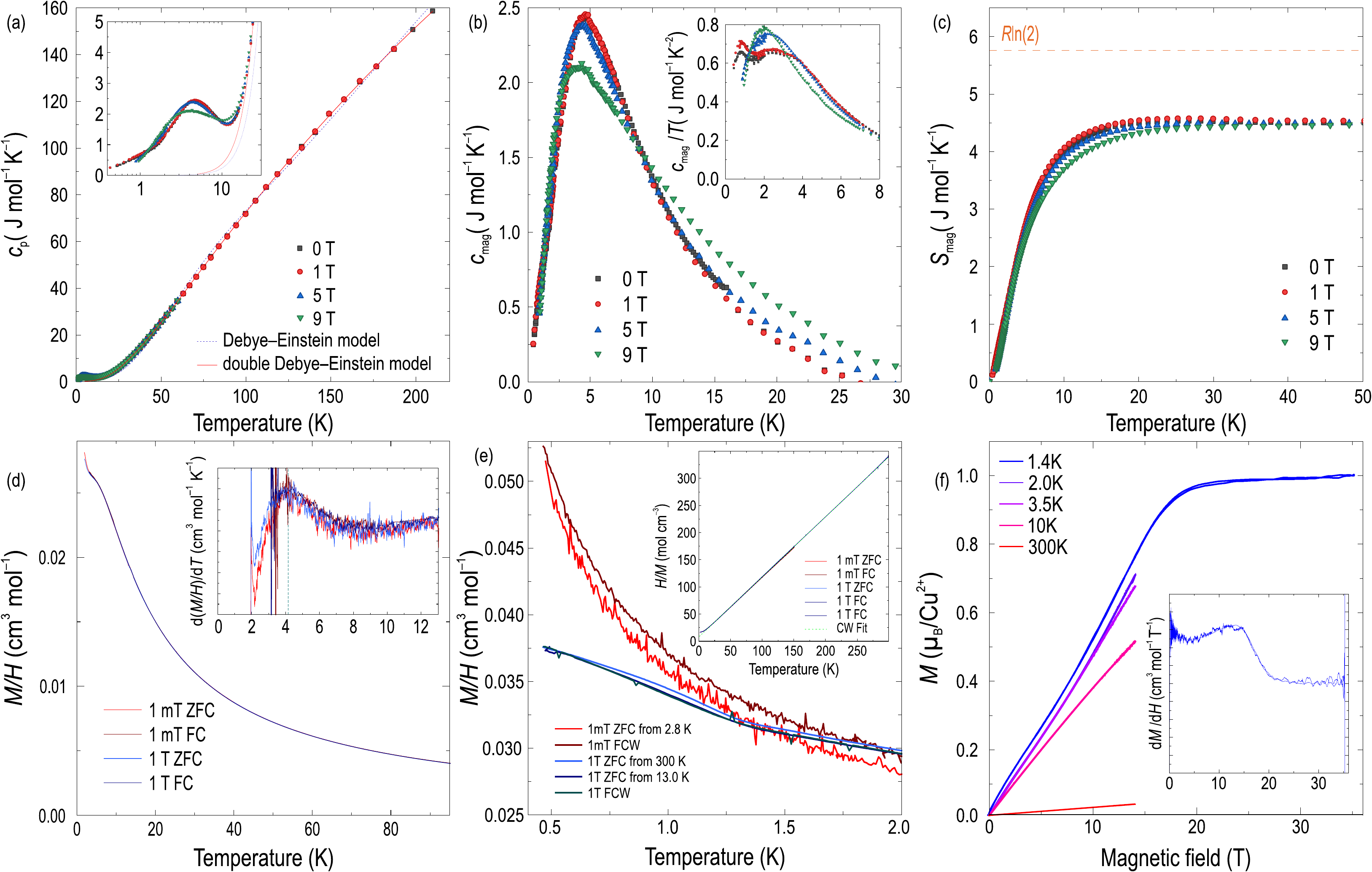}
		\caption{(a) Specific heat at various magnetic fields. Inset: low-temperature region. Blue dashed and red lines are fits to the single and double Debye-Einstein models, respectively. (b) Magnetic contribution to the specific heat at various magnetic fields. Inset: $c_\text{mag}/T$. (c) Magnetic entropy compared against the expected $R\ln2$. (d) Magnetic susceptibility. Inset: first derivative of the magnetic susceptibility. (e) Low-temperature magnetic susceptibility measured with a $^3$He refrigerator insert. Inset: inverse magnetic susceptibility. (f) Field-dependent magnetization at several temperatures. Inset: first derivative of the magnetization at 1.4\,K.}
		\label{fig:Cp_M}
	\end{figure*}
        
	Low-temperature specific-heat measurements were performed by the relaxation time method on a thin cold-pressed pellet of \cusnoh\ using a Quantum Design Physical Property Measurement System~(PPMS) both using a standard $^4$He specific heat puck and a respective puck in a $^3$He refrigerator. The sample was mounted to the measurement platform using Apiezon N grease; contributions from the sample holder and grease were subtracted. Multiple data points were collected at each temperature and averaged.
	
	Magnetization measurements down to 2\,K were performed by vibrating sample magnetometry~(VSM) in a Cryogenic Ltd.\ Cryogen-Free Measurement System~(CFMS), under zero-field-cooled-warming and field-cooled conditions.  Four-quadrant $M$--$H$ loops were measured to $\pm$14\,T at several temperatures.  Samples were loaded inside gel capsules which were inserted into plastic straws. Additional magnetization measurements down to 0.45\,K were performed in a Quantum Design Magnetic Property Measurement System~(MPMS-XL) equipped with an iHelium3 $^3$He refrigerator insert.

High-field magnetization measurements up to 35\,T at 1.4\,K were conducted at the Hochfeld-Magnetlabor Dresden~(HLD), Helmholtz-Zentrum Dresden-Rossendorf~(HZDR), in Dresden, Germany, using a pulsed magnet with a rise time of 7\,ms and a total pulse duration of 25\,ms. The magnetization was obtained by integrating the voltage induced in a compensated coil system surrounding the sample\,\cite{Skourski2011}.

\subsection{Neutron powder diffraction}
\label{SubSec:PND}

Neutron powder diffraction (NPD) measurements were conducted on \cusnod\ using the HRPT diffractometer\,\cite{HRPT_PSI} at SINQ, PSI, in Villigen, Switzerland, using neutrons of wavelength 2.450\,\AA. The counting times were approximately 26~hours each at 1.6 and 6\,K. 
    
    Additional NPD measurements were performed with 2.526-\AA\ neutrons on the D1B diffractometer\,\cite{Orench2014} at the Institut Laue-Langevin (ILL) in Grenoble, France, using a dilution refrigerator. The counting times were approximately 8 hours at 0.05\,K and 2 hours at 10\,K. The wavelength was selected using the (002) reflection from pyrolytic graphite.

\section{Results and Discussion}

\begin{figure}[tb]
		\includegraphics[width=\columnwidth]{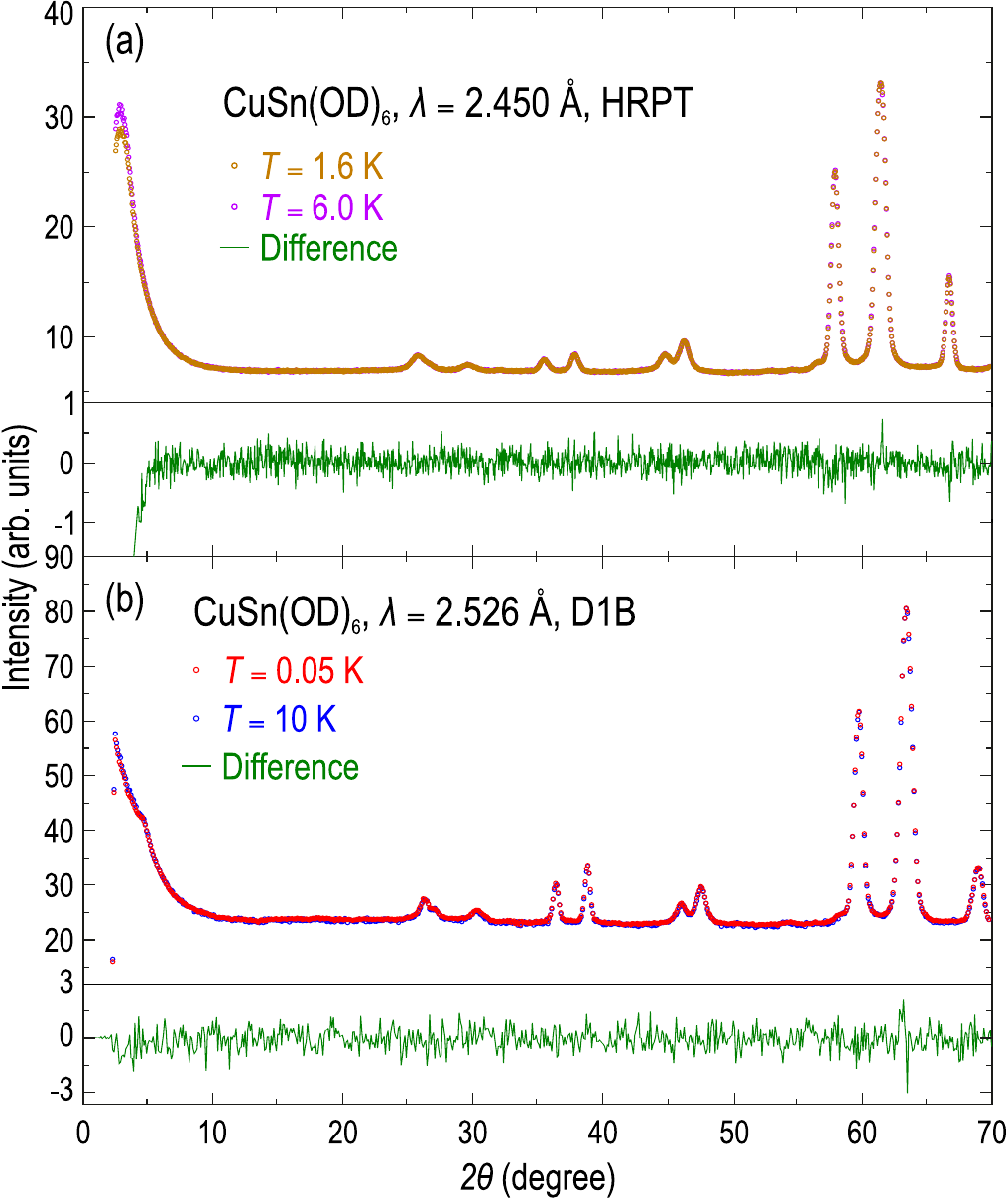}
		\caption{Neutron powder diffraction data from (a) HRPT at 1.6 and 6.0\,K and (b) D1B at 0.05 and 10\,K, and their differences, as a function of the scattering angle, $2\theta$.}
		\label{fig:neutrons}
	\end{figure}
	
The specific heat $c_P$ of \cusnoh\ is shown in Fig.\,\ref{fig:Cp_M}\,(a). Based on its color the material is assumed to be electrically insulating, so this can be described as $c_\text{p}(T)=c_\text{lattice}+c_\text{mag}$. The lattice contribution is typically modeled using Debye, Einstein, or a combination of both terms. However, in this case, a phenomenological function including two Debye and two Einstein terms is used:
\begin{equation}
\label{eq:clattice}
    c_\text{lattice}=m_{1}c_\text{D1}+(1-m_{1})c_\text{E1}+m_{2}c_\text{D2}+(1-m_{2})c_\text{E2}.
\end{equation} 
The contribution of the Debye heat capacity terms is
\begin{equation}
    c_\text{D}(T)=9Nk_\text{B} \left(\frac{T}{\theta_\text{D}}\right)^3 \int_{0}^{\theta_\text{D}/T} \frac{x^4 e^{x}}{\left(e^{x}-1\right)^{2}} \text{d}x,
\end{equation}
where $N$ is the number of participating atoms in the solid, $k_\text{B}$ is the Boltzmann constant and $\theta_{D}$ is the Debye temperature corresponding to the acoustic phonon vibrations in the sample. The contribution of the Einstein heat capacity terms at a constant volume is
\begin{equation}
    c_\text{E}(T)=3Nk_{B} \left(\frac{\theta_\text{E}}{T}\right)^{2} \frac{e^{\theta_\text{E}/T}}{\left(e^{\theta_\text{E}/T}-1\right)^2}
\end{equation}
where $\theta_\text{E}$ is the Einstein temperature corresponding to the optical phonons.  

Figure~\ref{fig:Cp_M}\,(a) shows results of the fit in the temperature range of 20--210 K using the single and double Debye-Einstein models, i.e., with two and four contributions, respectively. The double Debye-Einstein model shows significantly better agreement with the experimental data. %For this model, the fitted Debye and Einstein contributions  were $c_\text{D1}=48\,\%$, $c_\text{E1}=52\,\%$, $c_\text{D2}=21\,\%$~and~$c_\text{E2}=79\,\%$. The corresponding Debye and Einstein temperatures are $\theta^\text{D1}=1114(70)$, $\theta^\text{E1}=1748(700)$, $\theta^\text{D2}=212(5)$ and $\theta^\text{E2}=355(7)$~K. 

The double Debye-Einstein model fit was subtracted from the specific-heat data over the entire temperature range to obtain the magnetic contribution to the specific heat $c_\text{mag}$, shown in Fig.~\ref{fig:Cp_M}\,(b). The broad magnetic hump at 4.1\,K suggests the presence of short-range correlations. The low-temperature behavior follows neither a power-law dependence as $c_\text{mag} \sim T^\alpha$ with $\alpha = 1$ or 2, characteristic of gapless QSLs, nor an exponential as $c_\text{mag} \sim \exp(-\Delta/T$), characteristic of gapped QSLs. The hump also cannot be described as a Schottky-like anomaly. This feature broadens and shifts to lower temperature when a magnetic field is applied. The plot of $c_\text{mag}/T$ in the inset of Fig.~\ref{fig:Cp_M}\,(b) shows an additional magnetic-field-sensitive anomaly at about 0.8\,K, which is barely visible in $c_\text{mag}$.  The entropy in this peak is extremely small.  As shown below, this transition does not seem to be associated with long-range order.  

The magnetic entropy $S_\text{mag}$ obtained by integrating $c_\text{mag}/T$ is shown in Fig.~\ref{fig:Cp_M}\,(c). A point at (0,0) was added to all $c_\text{mag}/T$ curves for an approximate extrapolation to $T=0$; this extrapolated region accounts for $\sim$0.13\,J\,mol$^{-1}$\,K$^{-1}$.  The magnetic entropy saturates around 30\,K, far above the specific-heat anomalies as expected for frustrated spin systems, but it asymptotes only 85\% of $R\ln(2)$. This suggests that not all spin degrees of freedom are available at low temperatures. This may indicate residual quantum fluctuations, a magnetic transition at temperatures lower than our minimum of 0.45\,K, a model subtraction issue, or quantum entanglement suppressing classical thermal excitations. 

As shown in Fig.\,\ref{fig:Cp_M}\,(d), the magnetic susceptibility data also show a broad anomaly, which appears as a broad peak in the first derivative with a maximum at 4.1 K. No further significant features are observed down to 0.45\,K in low field, as shown in Fig.\,\ref{fig:Cp_M}\,(e). A subtle crossover appears in our magnetization data collected at 1\,T, where a slight kink appears at approximately 1.3\,K. However, due to the small magnitude of these changes, it is challenging to determine the nature of this feature. Fig.~\ref{fig:Cp_M}(e) shows both field-cooled and zero-field-cooled magnetization data, and no significant differences are found between these, indicating the absence of spin-glass behavior or any freezing of the spins. The Curie-Weiss temperature was obtained from a fit of $\chi^{-1}$ as shown in the inset of Fig.\,\ref{fig:Cp_M}\,(e). The estimated Curie-Weiss temperature $\Theta_\text{CW}=-7.1(3)$\,K, indicating dominant antiferromagnetic interactions, while the paramagnetic moment is $1.817(5)\,\mu_\text{B}$, slightly higher than the expected value for $S=\frac12$. The apparent lack of long-range magnetic order down to temperatures more than an order of magnitude below the Curie-Weiss temperature suggests strong frustration.  

Fig.~\ref{fig:Cp_M}\,(f) shows the magnetic field dependence of the magnetization at several temperatures, and up to 35\,T at 1.4\,K. An apparent saturation near 1\,$\mu_\text{B}$ per Cu is reached at roughly 18\,T. The 18-T energy scale exceeds that of the Curie-Weiss temperature by nearly a factor of two, suggesting a complex frustrated network of significant ferromagnetic and antiferromagnetic interactions which conspire to reduce the Curie-Weiss temperature.

To check for signs of magnetic order across the potential transitions at 0.8 and 4.1\,K, we performed NPD measurements at 0.05 and 10\,K at D1B, ILL, and at 1.6 and 6.0\,K on HRPT, at PSI (see Fig.~\ref{fig:neutrons}). We have already reported the crystal structure in detail elsewhere\,\cite{Kulbakov_2025}.  No features associated with magnetic long-range\,---\,or even short-range\,---\,order are seen down to 50\,mK, and no significant change in the background was observed at any angle.  This is in stark contrast to the sister compound MnSn(OD)$_\text{6}$, in which broad magnetic peaks were observed at low temperatures, associated with propagation vectors of $(\frac12\,\frac12\,\frac12)$ and/or $(0\,0.625\,0)$ but having a correlation length of only a few unit cells\,\cite{Parui2025Mn}.  We note that the emergence of order on the hydrogen site would be expected to leave strong signatures in the neutron diffraction pattern, so this can also be excluded.

Taking all our data together, we see no evidence of even short-range magnetic order down to 50\,mK, despite a Curie-Weiss temperature of $-$7.1\,K and an apparent interaction energy scale of 18\,T.  The nearly 2.5 orders of magnitude separating the temperature of our neutron measurement from the interaction energy scale suggests very strong frustration in \cusnod.  However, the suppression of magnetic order may also arise from the strong quantum fluctuations expected in $S=\frac{\text{1}}{\text{2}}$ Cu$^\text{2+}$ materials.  The Mn compound has a similar but higher-symmetric crystal structure with analogous proton disorder, but its magnetic moments are classical $S=\frac52$ spins which are not expected to experience significant quantum fluctuations.  The lower symmetry in the Cu analog would ordinarily be expected to partially relieve frustration, making long-range order more stable, so the lack of order suggests that quantum fluctuations play a key role in \cusnod.

Having established that \cusnod\ does not order down to 50\,mK, the question is what it does instead.  One possibility is that it is a quantum spin liquid.  We see evidence that quantum fluctuations may play an important role in the material, which would argue in favor of this scenario, but the strong Jahn-Teller distortions of the Cu coordination sphere lead to a rather low-symmetry crystal structure with a large number of inequivalent interactions, which would be expected to break the degeneracy required for a QSL ground state.  It is conceivable, although extremely unlikely, that these interactions are tuned to hit a pocket of QSL by sheer coincidence.  However, the correlated disorder on the hydrogen site adds another random fluctuation to the exchanges.  This makes a QSL extremely unlikely, in our opinion.  Instead, we propose that \cusnod\ exhibits spin liquid mimicry.  

The significant Zn/Cu cross-substitution in herbertsmithite\,\cite{Freedman_2010,Zorko_2017,Smaha_2020,Huang_2021,Imai_2011} leads to similar physics.  No long-range magnetic order is observed in herbertsmithite down to 50~mK despite strong antiferromagnetic interactions~\cite{Mendels_2007,Helton_2007,Misguich_2007, Chitra_2008}; however, the nature of the spin-liquid ground state remains under debate, and neutron and NMR data provide conflicting evidence. Some studies report a gapped~\cite{Imai_2011,Fu_2015,Han_2016} ground state while others suggest a gapless~\cite{Helton_2007,Khuntia_2020,Vries_2008} one.  Similarly, triangular-lattice YbMgGaO$_4$ with $j=\frac12$\,\cite{Li_2016} shows no magnetic order down to 50~mK, $\mu$SR measurements indicate persistent spin dynamics compatible with a U(1)-type QSL state, and inelastic neutron scattering further revealed a gapless continuum of magnetic excitations at 60~mK~\cite{Li_2019}. However, random site mixing between Mg$^{2+}$ and Ga$^{3+}$ leads to a locally disordered crystalline environment around Yb$^{3+}$ ions, resulting in spatially varying exchange interactions~\cite{Zhu_2017}. This structural randomness mimics spin-liquid behavior in YbMgGaO$_4$.

\begin{figure}[t]
    \includegraphics[width=\columnwidth]{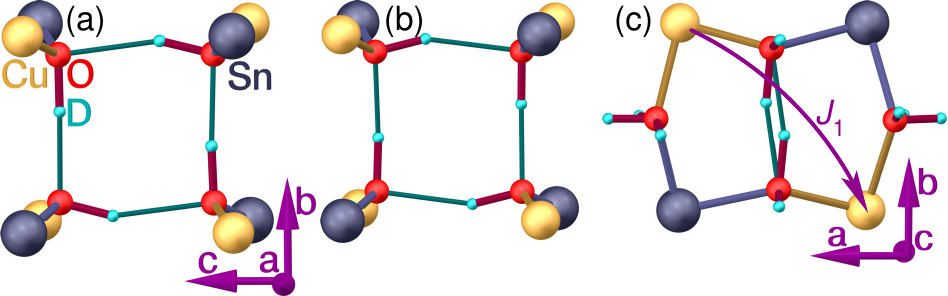}
    \caption{\label{exchanges}Role of hydrogen disorder on the Cu--Cu exchange pathways.  (a,b) Two possible configurations of the hydrogen atoms within a perovskite A site.  (c) Example of a $J_1$ nearest-neighbor exchange, with all (partially-occupied) hydrogen sites shown.}
\end{figure}

In both herbertsmithite and YbMgGaO$_4$, spin-liquid mimicry arises from quenched substitutional disorder that is difficult to control. In \cusnod, however, the disorder arises from ice-rule physics among the hydrogen atoms which fill the perovskite A site.  Figures~\ref{exchanges}(a) and \ref{exchanges}(b) show two possible hydrogen configurations in a selected A site.  These should be degenerate and completely decoupled from the other A sites.  An example nearest-neighbor exchange pathway $J_1$ is shown in Fig.~\ref{exchanges}(c), where all hydrogen positions are partially occupied.  Routes that do not pass through disordered hydrogen bonds traverse oxygen atoms to which disordered hydrogen atoms are bonded.  In water ice, hydrogen disorder can be tuned with pressure~\cite{Komatsu_2022}, leading to a rich phase diagram, and our previous neutron diffraction pressure study predicted that the hydrogen sites in \cusnod\ would merge at pressures on the order of 20\,GPa~\cite{Kulbakov_2025}, at which point the hydrogen disorder would be quenched. This is an experimentally-accessible pressure.   It is unlikely that all hydrogen sites would merge at the same pressure, potentially leading to a cascade of structural phases as seen in water ice.  It should be possible to tune the correlated hydrogen disorder out of existence while performing a variety of measurements, replacing the spin-liquid-like ground state we observe with either magnetic order or a quantum spin liquid. 

With magnetic order excluded as an explanation, and the ground state likely a spin liquid mimic, the hump in the specific heat at 4.1\,K may indicate gapless spin-liquid mimicry behavior, while the low-temperature anomaly at $\sim$0.8\,K may indicate the emergence of a spin gap. This would suggest that temperature-dependent magnetic exchange interactions could modify the excitation spectrum, leading to a crossover from gapless to possibly gapped spin-liquid mimicry. 

 \section{Conclusions}

\cusnoh\ does not order magnetically down to 50\,mK, and we find no evidence of spin freezing down to at least 0.45\,K.  Despite significant antiferromagnetic exchange interactions, a combination of quantum fluctuations, frustration, and hydrogen disorder prevents the formation of long-range magnetic order or any other static spin arrangement. Given the strong hydrogen disorder, we propose that \cusnoh\ is a rare example of spin-liquid mimicry, driven in this case by tunable structural disorder in the proton sublattice. In a unique difference from other such systems, however, in \cusnoh\ this disorder can be tuned.  The ability to control hydrogen disorder via external pressure~\cite{Kulbakov_2025} makes \cusnoh\ (and possibly some other hydroxide perovskites) a particularly promising platform for probing the interplay between quantum magnetism and structural disorder and exploring spin-liquid mimicry as a function of disorder strength.

\section*{Data Availability}
Samples and data are available upon reasonable request from D.~C. Peets or D.~S. Inosov; data underpinning this work is available from Refs.~\onlinecite{datasetCu2,5-31-2952}.

\begin{acknowledgments}
We gratefully acknowledge L.\ Zviagina for help with the pulsed-field magnetization experiments. This project was funded by the Deutsche Forschungsgemeinschaft (DFG, German Research Foundation) through: individual grants IN 209/12-1, DO 590/11-1 (Project No.\ 536621965), and PE~3318/2-1 (Project No.\ 452541981); through projects B01, B03, C01, and C03 of the Collaborative Research Center SFB~1143 (Project No.\ 247310070); and through the W\"urzburg-Dresden Cluster of Excellence on Complexity and Topology in Quantum Materials\,---\,\textit{ct.qmat} (EXC~2147, Project No.\ 390858490). The authors acknowledge the support of the Institut Laue-Langevin, Grenoble, France, as well as the HLD at HZDR, member of the European Magnetic Field Laboratory (EMFL). This work is based in part on experiments performed at the Swiss spallation neutron source SINQ, Paul Scherrer Institute, Villigen, Switzerland.
\end{acknowledgments}
	
%	\appendix
%	\newpage
	%\section{blau blau blau}\label{app:Exp}
	
	%blau blau bla
	
	\bibliography{CuSnOD6_mag}
	
\end{document}